\documentclass{aa}  

\usepackage{xcolor}
\usepackage[colorlinks=true, citecolor=blue]{hyperref}
\usepackage{graphicx}
\usepackage{txfonts}

\begin{document}

   \title{JESSNet: Joint dEconvolution and Sparse Separation Network \\ for HI 21 cm foreground cleaning}


   \author{Athanasia Gkogkou\inst{1}
          \and
          Victor Bonjean\inst{1}
          \and
          Jean-Luc Starck\inst{2,1}
          \and
          Pauline Gorbatchev\inst{1,4}
          \and
          Marta Spinelli\inst{5,6}
          \and
          Panagiotis Tsakalides\inst{3,7}
          }
           
   \institute{Institutes of Computer Science and Astrophysics, Foundation for Research and Technology Hellas (FORTH), Greece\\ \email{agkogkou@ia.forth.gr}
         \and
   Université Paris-Saclay, Université Paris Cité, CEA, CNRS, AIM, 91191, Gif-sur-Yvette, France
   \and
    Institute of Computer Science, Foundation for Research and Technology Hellas (FORTH), Greece
    \and
    Department of Physics, University of Crete, Greece
    \and
    Observatoire de la Côte d'Azur, Laboratoire Lagrange, Bd de l'Observatoire, CS 34229, F-06304 Nice Cedex 4, France
    \and
    Department of Physics and Astronomy, University of the Western Cape, Bellville, Cape Town 7535, South Africa
    \and
    Department of Computer Science, University of Crete, Greece
    }

   \date{Received 01/09/2026; Accepted 04/10/2026}

 
\abstract
{Foreground cleaning is one of the main challenges for H\,{\sc i} 21 cm intensity mapping, here considered over the redshift range
$z\simeq0.09\text{--}0.58$, especially in the presence of a chromatic instrumental beam that couples angular foreground structure into the frequency direction. This problem is further complicated by incomplete sky coverage, as expected for realistic Galactic masks and survey footprints.}
{We introduce JESSNet, a beam-aware and mask-compatible foreground-cleaning framework designed to preserve the statistical properties of the H\,{\sc i} signal in single-dish intensity mapping observations affected by frequency-dependent beam distortions and masked sky regions.}
{JESSNet advances the joint deconvolution and blind source-separation framework of DecGMCA and SDecGMCA through three methodological extensions. First, while the original formulation uses a single effective mixing model over the angular
range being cleaned, JESSNet operates in angular windows, with an effective mixing matrix and a channel selection adapted to each scale. Second, JESSNet replaces the fixed starlet thresholding step with a learned spherical-wavelet
"learnlet" sparse regularization operator. Third, whereas the original spherical implementation assumes full-sky coverage, JESSNet constrains the source estimates to the observed sky region, making the method applicable to Galactic
masks and survey footprints. We test the method on simulated SKA-MID-like observations covering $900\text{--}1300\,{\rm MHz}$, including H\,{\sc i} emission, Galactic and extragalactic foregrounds, thermal noise, and an oscillating in frequency MeerKAT-inspired beam model.}
{The multiscale JESSNet reconstruction accurately recovers the angular and frequency power spectra of the input H\,{\sc i} signal. Compared to applying JESSNet over a single angular range, the multiscale formulation significantly improves the reconstruction at low and intermediate angular scales and reduces residual beam-induced spectral features. The method also remains effective for a synthetic analysis footprint,
preserving the H\,{\sc i} power spectra under incomplete sky coverage.}
{JESSNet provides a beam-aware, multiscale foreground-cleaning framework for H\,{\sc i} intensity mapping. The code used to produce the analysis and results is publicly released, allowing the community to reproduce the pipeline and adapt it to future simulations and real observations.}


   \maketitle
%

\section{Introduction}
\label{sect:introduction}

Neutral hydrogen (H\,{\sc i}) intensity mapping has emerged as a powerful observational technique for mapping the large-scale structure of the Universe over wide areas and broad redshift ranges. Instead of detecting individual galaxies, H\,{\sc i} intensity mapping measures the integrated 21 cm emission from unresolved sources within large angular and frequency voxels \citep{battye2004,santos2005,chang2008,peterson2009,bernal2022, chang2026}. This makes it possible to probe cosmological volumes efficiently while retaining excellent radial information through the direct mapping between observed frequency and redshift. Early detections with the Green Bank Telescope \citep{chang2010,masui2013,switzer2013} and recent MeerKAT measurements, including calibrated single-dish autocorrelation maps and a significant cross-correlation with the WiggleZ galaxy survey, have demonstrated the observational feasibility of the technique \citep{wang2021,cunnington2023,cunnington2026}. Current and upcoming surveys with MeerKAT and SKA-Mid are therefore expected to play an important role in constraining the late-time matter distribution and the expansion history of the Universe \citep{santos2015,bull2015,bacon2020}.

A central challenge for H\,{\sc i} intensity mapping is the removal of astrophysical foregrounds. Galactic synchrotron emission, Galactic free-free emission, and extragalactic radio sources are several orders of magnitude brighter than the cosmological 21 cm signal \citep{santos2005,jelic2008,alonso2015,spinelli2021}. Although these foregrounds are expected to be spectrally smoother than the H\,{\sc i} signal, instrumental effects can strongly complicate this separation. In particular, the frequency-dependent telescope beam couples angular foreground structure into the frequency direction, producing artificial spectral fluctuations that can mimic or contaminate the cosmological signal \citep{shaw2015,matshawule2021,spinelli2021}. This effect is especially problematic when the beam does not vary smoothly with frequency, as is the case for realistic single-dish beam models with oscillatory features induced by the optical response of the telescope \citep{asad2021,matshawule2021,chen2025}.

A wide range of foreground-cleaning methods has been developed for H\,{\sc i} intensity mapping. Polynomial and parametric fitting methods exploit the expected spectral smoothness of the foregrounds, while blind source separation methods such as principal component analysis (PCA), independent component analysis (ICA), and generalized morphological component analysis (GMCA) aim to identify the dominant foreground subspace directly from the data \citep{liu2011,ansari2012,chapman2012,wolz2014,alonso2015,bigot-sazy2015,olivari2016,carucci2020,spinelli2021,gkogkou2026,spinelli2026}. These approaches have been shown to perform well in idealized cases, especially when all frequency channels are smoothed to a common angular resolution. However, their performance degrades when the beam is left at its native frequency-dependent resolution, because the standard linear mixture model no longer provides an accurate description of the data. Recent blind foreground-subtraction comparisons for SKA-MID- and MeerKAT-like simulations have further shown that interactions between foreground structure and instrumental beam effects can leave spurious artifacts in the cleaned maps, motivating foreground-cleaning methods that treat the beam response explicitly \citep{spinelli2021,gkogkou2026}.

A complementary, fully data-driven line of work applies deep neural networks directly to the observed maps \citep{makinen2021,chen2024}. These methods can achieve effective foreground suppression without an explicit physical mixing model, but their performance depends on the fidelity of the training simulations, and \citet{chen2024} showed that their robustness can likewise degrade under frequency-dependent instrumental systematics similar to those considered in this work.

To address this limitation, deconvolution and component separation must be treated jointly. DecGMCA introduced this idea in a sparse blind-source-separation framework \citep{jiang2017}, and SDecGMCA extended it to spherical data by incorporating the beam transfer function directly in harmonic space
\citep{carloni2021}. In \citet{gkogkou2026}, SDecGMCA was applied to simulated H\,{\sc i} intensity mapping observations and showed that it can recover the H\,{\sc i} signal more robustly than standard foreground-cleaning methods in the presence of frequency-dependent beam distortions. Nevertheless, a single global mixing model remains restrictive when the angular information available in the data changes strongly with frequency: broad-beam channels are informative on
large angular scales, but become increasingly ill-conditioned near their beam cut-off.

A complementary way to relax a global mixture model is to make the separation local in angular scale, or in both angular scale and position. This idea has been explored extensively in CMB component separation, for example through L-GMCA, where sparsity is exploited in a wavelet representation and the mixing model is allowed to vary across localized domains in order to account for beam differences and spatial variations of the foreground emission laws \citep{bobin2013}. Related scale-localized approaches have also recently been developed for H\,{\sc i} intensity mapping, including needlet-based PCA, needlet Karhunen--Loève cleaning, and needlet ILC/GNILC-type methods \citep{podczerwinski2024,dai2025,decaro2026}. These works support the
idea that the effective foreground subspace can depend on angular scale. In this work, we adopt this multiscale perspective within the SDecGMCA framework: rather than applying a scale-localized PCA or ILC alone, we retain the joint deconvolution and sparse source-separation formulation while allowing the effective mixing matrix and the retained channel set to vary across angular windows.

Two further limitations are important for applications to realistic data. First, sparsity-based methods depend on the representation in which the sources are assumed to be sparse. Classical wavelet dictionaries, such as starlets \citep{starck2007}, are interpretable and efficient, but they are not necessarily optimal for the complex morphology of astrophysical foregrounds. Learnlets provide a natural alternative: they are learned, wavelet-like multiscale transforms that retain
the structure and interpretability of sparse representations while adapting their filters and thresholds to the data \citep{ramzi2021}. Recently, \citet{bonjean2026} introduced the learnlet component separator, demonstrating that learnlets can improve blind source separation by replacing fixed sparse dictionaries with learned sparse operators. Second, real H\,{\sc i} surveys do
not observe a clean full sky. Galactic cuts, point-source masks, and survey footprints remove substantial sky regions, and any foreground-cleaning method intended for real observations must be able to operate under incomplete sky coverage.

In this paper we introduce JESSNet, the Joint dEconvolution and Sparse Separation Network\footnote{\url{https://github.com/Agkogkou/JESSNet}}. JESSNet builds on the SDecGMCA framework, preserving its joint treatment of beam deconvolution and blind source separation, but extends it in three directions. First, inspired by multiscale GMCA approaches, the foreground model is estimated in angular windows, with an effective mixing matrix and a set of frequency channels adapted to each angular scale. This allows the method to use low-resolution channels on large angular scales while avoiding their contribution to poorly resolved small-scale modes. Second, the fixed starlet thresholding step of SDecGMCA is replaced by a spherical-wavelet learnlet sparse regularization operator, which provides a data-adapted multiscale representation of the reconstructed sources. Third, the source estimates are constrained to the observed sky region, allowing the method to operate on Galactic masks and survey footprints.

We validate JESSNet on simulated single-dish H\,{\sc i} intensity mapping observations convolved with an oscillating MeerKAT-inspired beam model. We first test the method on full-sky simulations and compare the multiscale reconstruction to a JESSNet-only reconstruction without the angular-window
treatment. We then apply the method to a synthetic analysis footprint in order to assess its performance under incomplete sky coverage. The structure of this paper is as follows. In Sect.~\ref{sect:simulations}, we describe the simulated data. In Sect.~\ref{sect:jessnet}, we present the JESSNet framework, including the scale-dependent mixing model, angular-window channel selection, spherical-wavelet learnlet regularization, mask-constrained reconstruction, and
algorithmic implementation. In Sect.~\ref{sect:results}, we present the full-sky and masked-sky results. We summarize our findings in Sect.~\ref{sect:conclusions}.

\section{Simulations}
\label{sect:simulations}

We assess the performance of the component-separation pipeline using controlled simulations of single-dish HI intensity mapping observations. The simulated data cubes cover the frequency interval $900\text{--}1300\,{\rm MHz}$ ($z\simeq0.09\text{--}0.58$) with a frequency resolution of $\Delta \nu = 1 \, {\rm MHz}$ and contain the cosmological 21 cm signal, Galactic diffuse foregrounds, extragalactic radio sources, thermal noise, and the effect of a chromatic telescope beam. The sky maps are generated on the sphere using the HEALPix pixelization \citep{gorski2005} with ${\rm N}_{\rm side}=256$, corresponding to ${\rm N}_{\rm pix}=12{\rm N}_{\rm side}^2$ pixels per frequency channel.

For each observing frequency, the individual sky components are first produced at a common angular resolution and then added in temperature units. The resulting multifrequency sky is convolved with the instrumental beam described in Sect.~\ref{subsect:telescope_beams}, and a realization of Gaussian thermal noise is added independently in each channel. The simulated observation at frequency $\nu$ and sky position $p$ can therefore be written schematically as
\begin{equation}
    T_{\rm obs}(\nu,p) = \left[T_{\rm HI}(\nu,p) + T_{\rm fg}(\nu,p) \right] * B_\nu + T_{\rm N}(\nu,p),
    \label{eq:observed_temperature_simulations}
\end{equation}
where $T_{\rm HI}$ is the cosmological signal, $T_{\rm fg}$ denotes the sum of the foreground contributions, $B_\nu$ is the frequency-dependent beam, $*$ denotes convolution on the sphere, and $T_{\rm N}$ is the instrumental noise.

\subsection{HI cosmological signal}
\label{subsubsect:simulations_hi}

The cosmological 21 cm signal is generated with \texttt{CoLoRe}\footnote{\url{https://github.com/damonge/CoLoRe/tree/master}} \citep{ramirez-perez2022}, a public code designed to produce fast mock realizations of cosmological observables on the full-sky light cone. \texttt{CoLoRe} first generates a three-dimensional matter density field in a Cartesian box large enough to contain the requested light cone, and then projects the relevant tracer fields onto angular and redshift/frequency maps. We use the lognormal density-field prescription with a Cartesian grid of 
$N_{\rm grid}=2048$ cells per side to generate the three-dimensional matter density field. The adopted cosmological parameters are $\Omega_{\rm m}=0.3$, $\Omega_{\Lambda}=0.7$, $\Omega_{\rm b}=0.049$, 
$h=0.67$, $n_s=0.965$, and $\sigma_8=0.8$. We use \texttt{CAMB} to generate the input linear matter power spectrum at $z=0$, which is renormalized by \texttt{CoLoRe} to the value of $\sigma_8$.

For line-intensity mapping, \texttt{CoLoRe} treats the emitting species as a biased tracer of the underlying matter density field. The 21 cm brightness temperature in a direction $\hat{\mathbf{n}}$ and at observed frequency $\nu$ is written as
\begin{equation}
    T_{\rm HI}(\nu,\hat{\mathbf{n}}) = \overline{T}_{\rm HI}(z)
    \left[1+\delta_{\rm HI}(z,\hat{\mathbf{n}})\right],
    \label{eq:hi_brightness_temperature_colore}
\end{equation}
where $z=\nu_{21}/\nu-1$, $\nu_{21}=1420.405\,{\rm MHz}$ is the rest-frame frequency of the 21 cm transition, $\overline{T}_{\rm HI}(z)$ is the mean brightness temperature, and $\delta_{\rm HI}$ is the HI overdensity in 
redshift space. In our setup, the linear HI bias \citep{martin2012} and HI abundance \citep{crighton2015} evolve as
\begin{equation}
    b_{\rm HI}(z) = 0.3(1+z)+0.6,
    \label{eq:hi_bias_colore}
\end{equation}
and
\begin{equation}
    \Omega_{\rm HI}(z) = 4 \times 10^{-4}(1+z)^{0.6}.
    \label{eq:omega_hi_colore}
\end{equation}
The former sets the clustering bias of the HI field, while the latter determines the redshift evolution of the mean HI abundance and therefore the normalization of $\overline{T}_{\rm HI}(z)$.

To construct the intensity maps, \texttt{CoLoRe} assigns the contribution of the emitting volume elements to angular pixels and frequency bins. Redshift-space distortions are included by shifting the observed frequency 
according to the radial peculiar velocity of each emitting element.

\subsection{Astrophysical foregrounds}
\label{subsect:simulations_foregrounds}

The foreground model follows the same prescription as in \citet{gkogkou2026}. It includes Galactic synchrotron and free-free emission, generated with \texttt{PySM3} \citep{zonca2021}, and an extragalactic radio-source component based on the source-count model of \citet{battye2013}. The synchrotron emission is modeled as a power law in frequency with a spatially varying spectral index, using the reprocessed Haslam 408 MHz map \citep{remazeilles2015} as the amplitude template and the spectral-index map of \citet{miville2008}. The free-free emission is generated from the Planck Commander-based template and scaled with a fixed spectral index $\beta_{\rm ff}=-2.14$. The point-source contribution includes both clustered unresolved emission and randomly injected resolved sources, with sources brighter than $1\,{\rm Jy}$ assumed to be removed.

The frequency dependence of each foreground component is described by a power-law scaling,
\begin{equation}
    T_i(\nu,p) = T_i(\nu_0,p)
    \left(\frac{\nu}{\nu_0}\right)^{\beta_i(p)},
    \label{eq:foregrounds}
\end{equation}
where $T_i(\nu,p)$ is the brightness temperature of component $i$ at observing frequency $\nu$ and sky position $p$, $T_i(\nu_0,p)$ is the corresponding template at the reference frequency $\nu_0$, and $\beta_i(p)$ is the spectral index. For synchrotron and point sources, $\beta_i$ varies across the sky, while for free-free emission it is fixed to $\beta_{\rm ff}=-2.14$.

\subsection{Telescope beam}
\label{subsect:telescope_beams}

We model the instrumental response as an axisymmetric Gaussian beam with a frequency-dependent width. To capture departures from a purely smooth chromatic evolution, we adopt the MeerKAT-inspired parameterization of \citet{matshawule2021}, who showed that interactions between the primary and secondary reflectors introduce oscillatory features in the beam response. The beam profile is written as
\begin{equation}
    B_\nu(\theta) = \exp \left[-\frac{\theta^2}{2\sigma_\nu^2}\right],
    \label{eq:beam_response}
\end{equation}
where the Gaussian width is related to the full width at half maximum by $\sigma_\nu = \frac{{\rm FWHM}(\nu)}{\sqrt{8\ln 2}}$. The frequency dependence of the beam width is parameterized as
\begin{equation}
    {\rm FWHM}(\nu) = \frac{\lambda}{D} \left[\sum_{d=0}^{8} a_d \nu^d + A \sin \left(\frac{2\pi\nu}{T}  \right) \right],
    \label{eq:fwhm_oscillating}
\end{equation}
where $D=13.5\,{\rm m}$ is the dish diameter, $\lambda=c/\nu$, and the polynomial coefficients $a_d$ are taken from \citet{matshawule2021}. The sinusoidal term introduces a small oscillatory perturbation with period $T=20\,{\rm MHz}$ and amplitude $A=0.1\,{\rm arcmin}$. 

All frequency maps are convolved with this beam at their native resolution. We do not apply any additional smoothing to enforce a common angular resolution across frequencies, since the goal is to evaluate the ability of the method to handle the chromatic instrumental response directly.

\subsection{Instrumental noise}
\label{subsect:instrumental_noise}

The thermal noise is modeled as a Gaussian random field with zero mean, no spatial correlations, and no correlations between frequency channels. The noise variance is taken to be uniform across the observed sky at fixed frequency, but its amplitude changes with $\nu$ through the system temperature and beam solid angle. For each frequency channel, the noise standard deviation is
\begin{equation}
    \sigma_{\rm N}(\nu) = T_{\rm sys}(\nu) \left[ \frac{4\pi f_{\rm sky}} {\Delta\nu\, t_{\rm obs}\, N_{\rm dishes}\,\Omega_{\rm beam}(\nu)} \right]^{1/2},
\label{eq:sigma_instrumental_noise}
\end{equation}
where $f_{\rm sky}$ is the sky fraction, $\Delta\nu$ is the channel width, $t_{\rm obs}$ is the total integration time, $N_{\rm dishes}$ is the number of dishes, and $\Omega_{\rm beam}(\nu) =  1.133\,\theta_{\rm FWHM}^2(\nu)$ is the Gaussian beam solid angle. The system temperature includes an instrumental contribution and a sky-temperature term,
\begin{equation}
    T_{\rm sys}(\nu) =  T_{\rm instr} + 66 \left(\frac{300\,{\rm MHz}}{\nu} \right)^{2.55} {\rm K},
    \label{eq:temperature_instrumental_noise}
\end{equation}
where $\nu$ is expressed in MHz. We adopt survey parameters representative of SKA-MID-like single-dish observation, following \citet{santos2015}. We thus set $\Delta\nu = 1 \, {\rm MHz}, \, t_{\rm obs} = 20,000 \, {\rm h}, \, T_{\rm instr} = 25 \, {\rm K}, \, N_{\rm dishes} = 197$. We set $f_{\rm sky}=1$ for the full-sky analyses, while for the masked-sky cases we adjust $f_{\rm sky}$ to the unmasked sky fraction of the corresponding mask.

\section{JESSNet: Joint dEconvolution and Sparse Separation Network}
\label{sect:jessnet}

In this section, we introduce JESSNet, the Joint dEconvolution and Sparse Separation Network. JESSNet is a beam-aware foreground-cleaning framework that builds on DecGMCA \citep{jiang2017} and its spherical extension SDecGMCA \citep{carloni2021}. The method preserves the central principle of SDecGMCA, namely the joint treatment of deconvolution and blind source separation in harmonic space, but extends it in three directions. First, inspired by the multiscale philosophy of L-GMCA developed for CMB component separation \citep{bobin2013}, the reconstruction is embedded in a scale-dependent framework: the data are decomposed into angular windows, an effective mixing matrix is estimated independently at each scale, and only the frequency channels carrying reliable angular information are used in the corresponding separation. Second, the fixed starlet thresholding step of SDecGMCA is replaced by a spherical-wavelet learnlet sparse regularization operator \citep{ramzi2021,bonjean2026}. Third, the source estimates are explicitly constrained to the observed sky region, allowing the method to be applied to Galactic masks and survey footprints.

The multiscale aspect of JESSNet follows the same motivation as localized or multiscale GMCA methods: different angular scales do not necessarily contain the same effective foreground covariance, especially when the observations have different angular resolutions. Low-resolution channels are informative on large angular scales, but become unreliable close to the beam cut-off. Excluding such channels from the smallest angular scales avoids the noise amplification that would result from forcing all observations to a common high-resolution description. Unlike approaches that first equalize the channel resolutions and then perform independent component separation at each scale, JESSNet retains the DecGMCA/SDecGMCA principle of solving deconvolution and blind source separation jointly. Each angular window is therefore modeled with its corresponding channel-dependent beam operators and an effective scale-dependent mixing matrix.

The purpose of JESSNet is foreground cleaning rather than strict physical component recovery. The components reconstructed by the method should therefore be interpreted as effective latent components spanning the foreground subspace, rather than as direct estimates of individual astrophysical emission mechanisms such as synchrotron, free-free emission, and point sources. This distinction is important because the relevant objective is not to recover each foreground component individually, but to estimate the total foreground contribution sufficiently well that it can be subtracted while preserving the statistical properties of the H\,{\sc i} signal.

\subsection{Input preprocessing}

Before foreground separation, we apply a preprocessing step to reduce the influence of localized high-amplitude H\,{\sc i} peaks on the estimation of the foreground subspace. These features arise from rare, strongly overdense structures in the simulated cosmological H\,{\sc i} field. Along some lines of sight, they produce sharp enhancements over a small number of frequency channels which, when superimposed on the much smoother foreground emission, appear as localized bumps in the total temperature spectrum. The same effect was identified in \citet{gkogkou2026}, where such features were shown to originate from bright H\,{\sc i} structures rather than from the simulated extragalactic radio-source population. Such features can dominate the covariance of the data and therefore bias the estimation of the mixing matrix, even when they occupy only a small fraction of the data cube.

The filtering is performed independently along each line of sight. For every HEALPix pixel, the frequency spectrum is decomposed into four starlet detail scales and one coarse scale. The localized H\,{\sc i} peaks are predominantly captured by the finest spectral scales. A robust dispersion estimate is computed from the wavelet coefficients using the median absolute deviation (MAD), and large positive coefficients exceeding a threshold of $\sigma_{\rm mask}$ times this dispersion are set to zero on the two finest detail scales. The filtered spectrum is then reconstructed by summing the modified wavelet coefficients.

The resulting filtered data cube is used only for estimating the foreground model, including the PCA foreground subspace and the JESSNet source and mixing matrix estimates. The final foreground estimate is subtracted from the original, unfiltered observations. The preprocessing therefore prevents localized bright H\,{\sc i} peaks from biasing the estimation of the smooth foreground subspace, without directly removing these features from the final reconstructed H\,{\sc i} data cube.

\subsection{Scale-dependent mixing model}

Taking into account the beams of the different frequency channels, the standard linear blind source-separation model must be generalized to a convolutive mixture. At frequency $\nu$, the observed map can be written schematically as
\begin{equation}
    \mathbf{X}_{\nu}
    =
    \left(
    \sum_{n=1}^{N_s}
    A_{\nu n}\mathbf{S}_{n}
    \right)
    \ast
    \mathbf{H}_{\nu}
    +
    \mathbf{N}_{\nu},
    \label{eq:jessnet_data_model}
\end{equation}
where $\mathbf{X}_{\nu}$ is the observed map, $\mathbf{S}_{n}$ is the $n$th source map, $A_{\nu n}$ is the mixing coefficient of source $n$ at frequency $\nu$, $\mathbf{H}_{\nu}$ is the frequency-dependent beam operator, $\mathbf{N}_{\nu}$ is the noise, and $N_s$ is the number of reconstructed components.

For an axisymmetric beam, the convolution is diagonal in spherical harmonic space. The model can therefore be written as
\begin{equation}
    \hat{\mathbf{X}}_{\nu}^{\ell m}
    =
    \hat{H}_{\nu}^{\ell}
    \sum_{n=1}^{N_s}
    A_{\nu n}
    \hat{\mathbf{S}}_{n}^{\ell m}
    +
    \hat{\mathbf{N}}_{\nu}^{\ell m},
    \label{eq:jessnet_harmonic_model}
\end{equation}
where $\ell$ and $m$ denote spherical harmonic indices, hats indicate harmonic-space quantities, and $\hat{H}_{\nu}^{\ell}$ is the beam transfer function of channel $\nu$.

In JESSNet, this model is extended to a multiscale setting. We introduce a set of angular windows $W_j(\ell)$, with $j=1,\dots,N_j$, and define the windowed harmonic data as
\begin{equation}
    \hat{\mathbf{X}}_{\nu,\ell m}^{(j)}
    =
    W_j(\ell)\,
    \hat{\mathbf{X}}_{\nu,\ell m}.
    \label{eq:jessnet_windowed_data}
\end{equation}
The formulation is written for a generic set of angular windows. The only requirement is that they define the angular scales over which independent foreground models are estimated and that their combination provides a well-defined reconstruction over the retained multipole range.

For each angular window, we introduce an effective scale-dependent mixing matrix $\mathbf{A}^{(j)}$. For the subset of channels retained at scale $j$, denoted by $\mathcal{C}_j$, the model becomes
\begin{equation}
    \hat{\mathbf{x}}_{\ell m}^{(j)}
    =
    \mathbf{H}_{\ell}^{(j)}
    \mathbf{A}^{(j)}
    \hat{\mathbf{s}}_{\ell m}^{(j)}
    +
    \hat{\mathbf{n}}_{\ell m}^{(j)} ,
    \label{eq:jessnet_scale_dependent_model}
\end{equation}
where $\hat{\mathbf{x}}_{\ell m}^{(j)}$ is the vector of windowed observations over the channels $\nu\in\mathcal{C}_j$, $\mathbf{H}_{\ell}^{(j)}$ is a diagonal matrix containing the beam transfer functions $\hat{H}_{\nu}^{\ell}$ for the retained channels, $\mathbf{A}^{(j)}$ is the effective mixing matrix at angular window $j$, and $\hat{\mathbf{s}}_{\ell m}^{(j)}$ are the corresponding source coefficients.

The scale dependence of $\mathbf{A}^{(j)}$ should be understood as an algorithmic relaxation rather than as a physical property of the sky. In the absence of instrumental effects, the sky foregrounds may be described by a single mixing model. However, after convolution with a frequency-dependent beam, masking, noise contamination, and channel selection, the effective foreground covariance becomes scale dependent. Allowing the mixing matrix to vary across angular windows gives the inversion additional flexibility and improves the approximation of the total foreground subspace in a highly ill-conditioned problem.

For each angular window, JESSNet alternates between a beam-aware source update, a learned sparse regularization step, and a mixing-matrix update. The corresponding problem can be written schematically as
\begin{equation}
    \min_{\mathbf{A}^{(j)},\,\mathbf{S}^{(j)}}
    \frac{1}{2}
    \left\|
    \hat{\mathbf{X}}^{(j)}
    -
    \hat{\mathbf{H}}^{(j)}
    \mathbf{A}^{(j)}
    \hat{\mathbf{S}}^{(j)}
    \right\|_{\rm F}^{2}
    +
    \lambda\,
    \mathcal{R}_{\theta}
    \left(
    \mathbf{S}^{(j)}
    \right),
    \label{eq:jessnet_learned_regularized_problem}
\end{equation}
where $\mathcal{R}_{\theta}$ denotes the learned sparse regularization induced by the spherical-wavelet learnlet operator. In practice, we do not explicitly minimize a closed-form analytic penalty $\mathcal{R}_{\theta}$. Instead, the regularization step is implemented as a learned proximal-like operator applied to the current source estimates. To remove the scale degeneracy inherent to blind source separation, the columns of the mixing matrix are normalized,
\begin{equation}
    \left\|
    \mathbf{A}^{(j)}_{\cdot n}
    \right\|_2 = 1,
    \qquad
    n=1,\dots,N_s .
    \label{eq:jessnet_column_normalization}
\end{equation}

\subsection{Angular windows and channel selection}

In the applications presented in this work, the angular windows $W_j(\ell)$ are chosen to be cosine-tapered windows. These windows preserve a set of user-defined multipole boundaries while replacing sharp transitions by local overlaps. For a boundary $\ell_b$ and taper width $\Delta\ell$, the transition is controlled by the smooth-step function
\begin{equation}
    q(x)
    =
    \frac{1}{2}
    \left[
    1-\cos(\pi x)
    \right],
    \qquad
    0\leq x \leq 1,
    \label{eq:jessnet_cosine_smoothstep}
\end{equation}
with $q(x)=0$ for $x<0$ and $q(x)=1$ for $x>1$. Around each boundary, neighboring windows are tapered over the interval $\ell_b-\Delta\ell \leq \ell \leq \ell_b+\Delta\ell$. After construction, the windows are normalized so that
\begin{equation}
    \sum_{j=1}^{N_j} W_j(\ell) = 1
    \label{eq:jessnet_window_partition}
\end{equation}
wherever the reconstruction has support. This partition-of-unity normalization ensures that the final recombination of the cleaned windowed contributions preserves the overall harmonic amplitude and avoids artificial discontinuities between adjacent angular ranges.

The channel selection is defined from the beam transfer functions, which are assumed to be known from instrumental characterization or from a forward model of the telescope response. For each angular window $W_j(\ell)$, we retain only the channels whose beam response remains sufficiently large over the multipole range probed by that window. Since the windows are smooth and overlapping, the selection is not defined from a single multipole boundary. Instead, we compute a window-weighted beam response,
\begin{equation}
    \overline{B}_{\nu}^{(j)}
    =
    \frac{
    \sum_{\ell} W_j^2(\ell)\,B_{\nu}(\ell)
    }{
    \sum_{\ell} W_j^2(\ell)
    },
    \label{eq:jessnet_weighted_beam_response}
\end{equation}
and define the retained channel set as
\begin{equation}
    \mathcal{C}_j
    =
    \left\{
    \nu :
    \overline{B}_{\nu}^{(j)}
    \geq
    \tau
    \right\}.
    \label{eq:jessnet_channel_selection}
\end{equation}
As the angular windows move to higher multipoles, progressively lower-resolution channels are removed, and only the channels that still carry reliable angular information at the corresponding scales are kept.

The lowest angular window is treated separately. On these large angular scales, all frequency channels retain useful information and the foregrounds remain highly correlated across frequency. We therefore clean the lowest angular window using PCA after homogenizing the angular resolution of the channels to a common reference beam. For all higher angular windows, we apply JESSNet using the channel set $\mathcal{C}_j$ selected by Eq.~\ref{eq:jessnet_channel_selection}.

After each angular window has been cleaned, the final reconstructed
H\,{\sc i} harmonic cube is obtained by recombining the cleaned estimates
with the same angular windows used in the analysis,
\begin{equation}
    \widehat{\rm HI}_{\nu,\ell m}
    =
    \frac{
    \sum_{j=1}^{N_j}
    W_j(\ell)\,
    \widehat{\rm HI}^{(j)}_{\nu,\ell m}
    }{
    \sum_{j=1}^{N_j}
    W_j(\ell)
    } .
    \label{eq:jessnet_multiscale_recombination}
\end{equation}
Here $\widehat{\rm HI}^{(j)}_{\nu,\ell m}$ denotes the cleaned estimate
obtained from the $j$th angular window before the final synthesis weighting.
In this work, the cosine-tapered windows are normalized to form a partition
of unity over the retained multipole range, so that
$\sum_j W_j(\ell)=1$ wherever the reconstruction has support. Equation
\ref{eq:jessnet_multiscale_recombination} therefore reduces in practice to
a weighted summation of the cleaned window contributions.

\subsection{Alternating minimization}

For each angular window $j$, JESSNet estimates the source coefficients and the effective mixing matrix through an alternating-minimization procedure. At fixed mixing matrix $\mathbf{A}^{(j)}$, the source coefficients are updated by solving a beam-aware deconvolution problem in harmonic space, followed by the spherical-wavelet learnlet regularization described in Sect.~\ref{subsect:spherical_learnlets}. At fixed regularized sources, the mixing matrix is updated by least squares using the detail-scale harmonic coefficients of the data and sources. These two steps are alternated during a warm-up phase and a refinement phase until convergence.

\subsubsection{Source update for fixed mixing matrix}

For a fixed angular window $j$ and a fixed mixing matrix $\mathbf{A}^{(j)}$, the source coefficients are estimated through a Tikhonov-regularized beam-aware least-squares update at each harmonic mode. Using the compact notation of Eq.~\ref{eq:jessnet_scale_dependent_model}, this update reads 
\begin{equation}
    \hat{\mathbf{s}}_{\ell m}^{(j)}
    =
    \left[ \left(\mathbf{A}^{(j)}\right)^{\top} \left(\mathbf{H}_{\ell}^{(j)}\right)^2 \mathbf{A}^{(j)}
    + \boldsymbol{\epsilon}_{\ell}^{(j)} \right]^{-1} \left(\mathbf{A}^{(j)}\right)^{\top} \mathbf{H}_{\ell}^{(j)}
    \hat{\mathbf{x}}_{\ell m}^{(j)},
    \label{eq:jessnet_source_update}
\end{equation}
where $\boldsymbol{\epsilon}_{\ell}^{(j)}$ is a diagonal regularization term.
As in SDecGMCA, we use the mixing-matrix-based regularization strategy of
\citet{carloni2021} to stabilize the inversion at multipoles where the
deconvolution becomes ill-conditioned.

The updated source coefficients are transformed back to map space and
regularized with the spherical-wavelet learnlet operator. The regularized source
maps are then projected onto the observed sky region and transformed back to
spherical harmonics. Only their detail-scale coefficients are retained for the
subsequent update of the mixing matrix.

\subsubsection{Mixing-matrix update for fixed sources}

For fixed regularized source estimates, the effective mixing matrix $\mathbf{A}^{(j)}$ is updated by least squares using the detail-scale harmonic coefficients of the data and sources. We denote these coefficients by $\hat{x}_{\nu,\ell m,{\rm det}}^{(j)}$ and $\hat{\mathbf{s}}_{\ell m,{\rm det}}^{(j)}$, respectively. For each retained frequency channel $\nu\in C_j$, the row vector $\mathbf{A}_{\nu}^{(j)}$ of the mixing matrix is obtained by solving
\begin{equation}
    \mathbf{A}_{\nu}^{(j)}
    =
    \arg\min_{\mathbf{A}_{\nu}^{(j)}} \sum_{\ell m} \left| \hat{x}_{\nu,\ell m,{\rm det}}^{(j)}
    -
    H_{\nu,\ell}^{(j)} \mathbf{A}_{\nu}^{(j)} \hat{\mathbf{s}}_{\ell m,{\rm det}}^{(j)} \right|^2.
    \label{eq:jessnet_mixing_update_objective}
\end{equation}
The corresponding least-squares solution is
\begin{equation}
    \mathbf{A}_{\nu}^{(j)}
    =
    \left( \sum_{\ell m} \hat{x}_{\nu,\ell m,{\rm det}}^{(j)} H_{\nu,\ell}^{(j)}
    \hat{\mathbf{s}}_{\ell m,{\rm det}}^{(j)\dagger} \right)
    \left( \sum_{\ell m} \left(H_{\nu,\ell}^{(j)}\right)^2 \hat{\mathbf{s}}_{\ell m,{\rm det}}^{(j)}
    \hat{\mathbf{s}}_{\ell m,{\rm det}}^{(j)\dagger} \right)^{-1},
    \label{eq:jessnet_mixing_update}
\end{equation}
where $\dagger$ denotes Hermitian transpose. This update is the beam-aware least-squares estimate of the mixing coefficients for channel $\nu$, analogous to the SDecGMCA update of \citet{carloni2021}, but applied independently in each angular window.

After the least-squares update, the columns of $\mathbf{A}^{(j)}$ are normalized according to Eq.~\ref{eq:jessnet_column_normalization}, which fixes the scale degeneracy between the mixing matrix and the source amplitudes. Positivity of the mixing matrix can also be imposed when required.

\subsubsection{Convergence}

Convergence is monitored through the relative variation of the source estimates
between two consecutive iterations. For each angular window, we define
\begin{equation}
    \Delta_S^{(j,t)}
    =
    \frac{
    \left\|
    \left(\mathbf{S}^{M,(j)}\right)^{(t)}
    -
    \left(\mathbf{S}^{M,(j)}\right)^{(t-1)}
    \right\|_{\rm F}
    }{
    \left\|
    \left(\mathbf{S}^{M,(j)}\right)^{(t)}
    \right\|_{\rm F}
    },
    \label{eq:jessnet_convergence}
\end{equation}
where $\mathbf{S}^{M,(j)}=M\mathbf{S}^{(j)}$ denotes the source estimates
restricted to the observed sky region. The iterations are stopped when this
relative variation falls below the adopted tolerance, or when the maximum
number of iterations is reached.

\subsection{Spherical-wavelet learnlet sparse regularization}
\label{subsect:spherical_learnlets}

The learnlet transform is a learned, wavelet-like multiscale representation designed to preserve the interpretability and sparsity-promoting properties of classical wavelets while adapting the filters and thresholds to the signal morphology. In JESSNet, the learnlet module replaces the fixed starlet thresholding used in SDecGMCA. This choice is motivated by the learnlet component separator (LCS) introduced by \citet{bonjean2026}, which showed that replacing fixed sparse dictionaries by learned wavelet-like operators can improve blind source separation relative to classical methods such as GMCA. In this work, we extend this idea to spherical H\,{\sc i} intensity-mapping data by applying a spherical-wavelet learnlet regularization to HEALPix source maps inside the beam-aware JESSNet alternating minimization.

For each angular window $j$, the current source estimates are first converted from spherical harmonics to map space,
\begin{equation}
    \mathbf{S}^{(j)} = \mathcal{F}^{\dagger} \hat{\mathbf{S}}^{(j)},
    \label{eq:jessnet_inverse_harmonic}
\end{equation}
where $\mathbf{S}^{(j)}$ denotes the stack of source maps associated with angular window $j$, $\mathcal{F}^{\dagger}$ denotes the inverse spherical harmonic transform, and $s_i^{(j)}$ denotes its $i$th component, obtained by transforming the harmonic coefficients $\hat{s}_{i,\ell m}^{(j)}$ of Eq.~\ref{eq:jessnet_source_update} back to map space.

For each source component, the learnlet is applied to dimensionless fluctuations rather than to the raw source amplitude. This makes the thresholding strength comparable across source components and angular windows, and is consistent with the unit-variance normalization used during learnlet training. We therefore normalize each source map on the observed sky region,
\begin{equation}
    \mu_n^{(j)} = \left\langle s_n^{(j)}(p) \right\rangle_{M(p)\neq 0}, \qquad \eta_n^{(j)}
    =
    {\rm std} \left( \{s_n^{(j)}(p):M(p)\neq 0\} \right),
    \label{eq:jessnet_source_normalization_stats}
\end{equation}
and define
\begin{equation}
    s_{n,{\rm norm}}^{(j)}(p) = \frac{s_n^{(j)}(p)-\mu_n^{(j)}}{\eta_n^{(j)}}.
    \label{eq:jessnet_source_normalization}
\end{equation}
The normalized source map is decomposed using a spherical wavelet transform,
\begin{equation}
    \left\{w_{i,q}^{(j)}(p) \right\}_{q=0}^{L-1}
    =
    \Psi_{\rm sph} \left(s_{i,{\rm norm}}^{(j)} \right),
    \label{eq:jessnet_spherical_wavelet_transform}
\end{equation}
where $\Psi_{\rm sph}$ denotes the spherical wavelet transform, and $q=0,\dots,L-1$ indexes the $L$ spherical-wavelet scales. In this work, we use $L=5$, corresponding to one coarse scale and four detail scales. We use $q$ for these internal learnlet scales to distinguish them from the outer angular
windows indexed by $j$.

For a given patch index $r$, we extract the local patch centered at that position from each spherical-wavelet scale. These $L$ aligned scale patches are then stacked into a multiscale patch cube,
\begin{equation}
    \mathbf{x}_{n,r}^{(j)} = \left[ \mathcal{P}_{r} w_{n,0}^{(j)},
    \mathcal{P}_{r} w_{n,1}^{(j)}, \dots, \mathcal{P}_{r} w_{n,L-1}^{(j)}
    \right] \in  \mathbb{R}^{L\times N_{\rm p}\times N_{\rm p}},
    \label{eq:jessnet_multiscale_patch_cube}
\end{equation}
where $\mathcal{P}_{r}$ extracts the local patch at position $r$, and $N_{\rm p}$ is the linear patch size. Each learnlet input therefore contains the same sky region observed across all spherical-wavelet scales.

The noise-level parameter provided to the learnlet is estimated adaptively from the normalized source map,
\begin{equation}
    \sigma_n^{(j)}
    =
    {\rm MAD}
    \left(
    \{s_{n,{\rm norm}}^{(j)}(p):M(p)\neq 0\}
    \right).
    \label{eq:jessnet_learnlet_sigma}
\end{equation}
This scalar sets the overall noise scale of the current source estimate. The thresholding strength is then determined inside the trained learnlet: at spherical-wavelet scale $q$, the hard-thresholding level is proportional to $k_q \sigma_n^{(j)}$, where the coefficients $k_q$ are learned during training.

\begin{figure*}[t!]
    \centering
    \includegraphics[width=\textwidth]{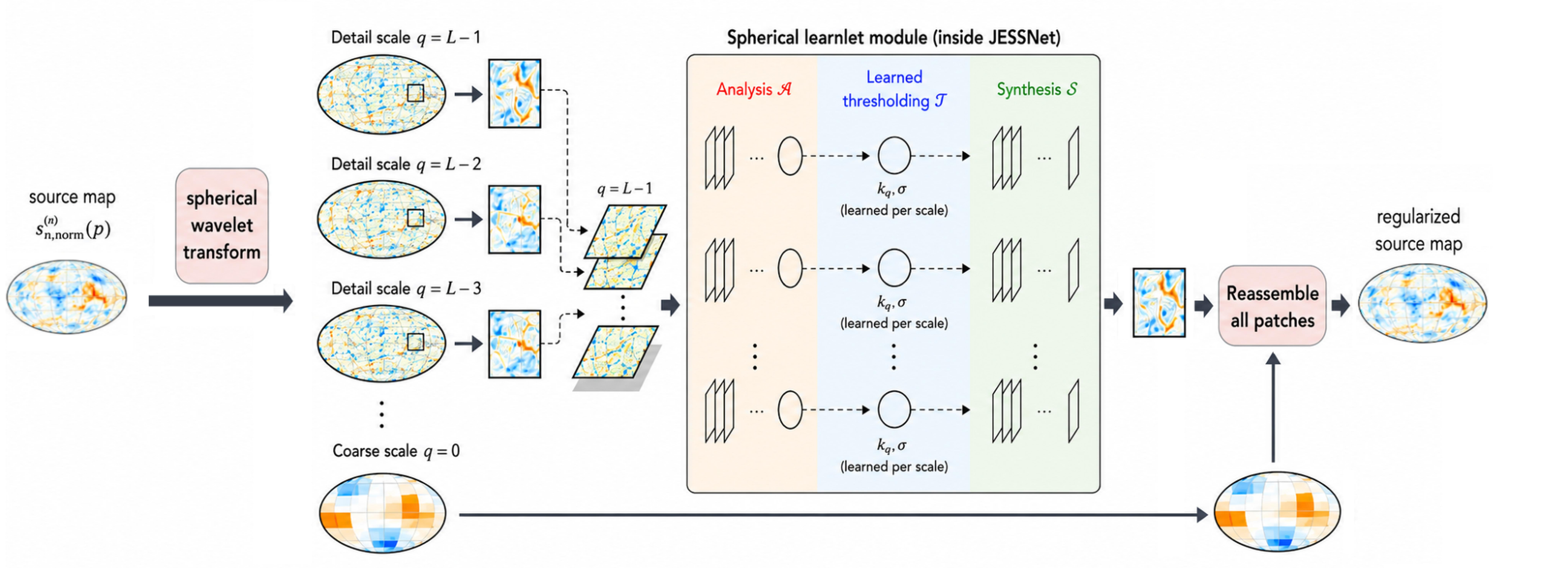}
    \caption{Schematic representation of the spherical-wavelet learnlet regularization used inside JESSNet, showing the construction of multiscale patch cubes from spherical-wavelet scales and their denoising through the learned analysis, thresholding, and synthesis operators.}
    \label{fig:spherical_learnlets}
\end{figure*}

The learnlet operator is applied to each multiscale patch cube,
\begin{equation}
    \widetilde{\mathbf{y}}_{n,r}^{(j)}
    =
    \mathcal{L}_{\theta}
    \left(
    \mathbf{x}_{n,r}^{(j)};
    \sigma_n^{(j)}
    \right),
    \label{eq:jessnet_spherical_learnlet_operator}
\end{equation}
where $\theta$ denotes the learned parameters of the spherical-wavelet learnlet, including the analysis filters, the synthesis filters, and the scale-dependent threshold coefficients. The output is a denoised map patch, rather than a set of denoised spherical-wavelet coefficients. The denoised patches are reassembled into a full normalized HEALPix map and the source map is then returned to its original normalization. Applying this operation to all source components gives the thresholded source matrix $\mathbf{S}_{\rm thr}^{(j)}$. The learnlet procedure is visualized in Fig.\,\ref{fig:spherical_learnlets}.

After learnlet regularization, the reconstructed source maps are converted back to spherical harmonics,
\begin{equation}
    \hat{\mathbf{S}}_{\rm thr}^{(j)} = \mathcal{F} \left( \mathbf{S}_{\rm thr}^{(j)} \right),
    \label{eq:jessnet_forward_harmonic}
\end{equation}
and only their detail scales are used in the subsequent update of the mixing matrix.

\subsection{Training of the learnlet}
\label{subsect:learnlet_training}

The learnlet is trained on simulated maps of the individual foreground components used in our analysis, namely Galactic synchrotron, Galactic free-free emission, and extragalactic radio sources. We construct a set of spectral variants around the fiducial foreground prescriptions adopted in Sect.~\ref{subsect:simulations_foregrounds}. For the point-source component, we vary the mean spectral index over $\beta_{\rm ps}=\{-2.5,-2.6,-2.7,-2.8,-2.9\}$, while retaining a spatial scatter of $\sigma_{\beta}=0.2$. For Galactic synchrotron, the spectral-index map of the \texttt{PySM3 s1} model is shifted by $\Delta\beta_{\rm sync}=\{-0.2,-0.1,0,0.1,0.2\}$, and for free-free emission we use $\beta_{\rm ff}=\{-2.1,-2.14,-2.2\}$. These values are intended to sample a physically plausible range around the fiducial spectral behavior rather than arbitrary spectral laws.

To train the learnlet as a denoising regularizer, Gaussian noise is added directly to the multiscale input patches. For each sample, the noise amplitude $\sigma$ is drawn independently in the interval $0\leq\sigma\leq\sigma_{\rm max}$, with $\sigma_{\rm max}=0.5$, and independent Gaussian noise is then generated at this amplitude. The network is trained to reconstruct the corresponding clean normalized map patch.

The spectral variants used for training are intended to make the learned regularization robust to moderate uncertainties in the foreground spectra while remaining within physically plausible ranges. If the true sky contains substantially different morphologies, spectral curvature, or other unmodeled complexity, the effectiveness of the learnlet regularization may degrade and subsequently affect the estimated foreground model. We emphasize, however, that the learnlet acts only as a sparse regularizer within JESSNet; the scale-dependent mixing matrices and source amplitudes are still estimated blindly from the observed data. The training strategy could be further improved by broadening the simulated foreground diversity, enabling a more systematic assessment of robustness to training-to-data mismatch.

\subsection{Mask-constrained reconstruction}

Real observations are affected by masks arising from survey boundaries, Galactic cuts, point-source removal, or other data-quality selections. Let $M(p)$ denote a binary mask, with $M(p)=1$ for observed pixels and $M(p)=0$ for excluded pixels. In the absence of masking, the observed map at frequency $\nu$ is modeled as
\begin{equation}
    \mathbf{X}_{\nu}
    =
    \left(
    \mathbf{A}_{\nu}\mathbf{S}
    \right)
    \ast
    \mathbf{H}_{\nu}
    +
    \mathbf{N}_{\nu}.
    \label{eq:jessnet_fullsky_mask_model}
\end{equation}
After applying a mask, the exact masked data are
\begin{equation}
    \mathbf{X}^{M}_{\nu}
    =
    M
    \left[
    \left(
    \mathbf{A}_{\nu}\mathbf{S}
    \right)
    \ast
    \mathbf{H}_{\nu}
    \right]
    +
    M\mathbf{N}_{\nu}.
    \label{eq:jessnet_exact_masked_model}
\end{equation}
This expression shows that the mask acts after beam convolution. In general, multiplication by the mask and convolution with the beam do not commute. The mask therefore introduces boundary effects in map space and mode coupling in harmonic space.

We do not attempt to invert the full mask-induced mode-coupling operator inside JESSNet. Instead, we make use of the fact that the mask affects the reconstruction most strongly near its boundaries. For pixels sufficiently far from the mask edge, the contribution from emission outside the observed region to the beam-convolved signal is expected to be subdominant. Under this assumption, the masked data model can be approximated by moving the mask inside the source term,
\begin{equation}
    \mathbf{X}^{M}_{\nu}
    \simeq
    \left[
    \mathbf{A}_{\nu}
    \left(
    M\mathbf{S}
    \right)
    \right]
    \ast
    \mathbf{H}_{\nu}
    +
    M\mathbf{N}_{\nu}.
    \label{eq:jessnet_approx_masked_model}
\end{equation}
This approximation is expected to be most accurate away from the mask boundary and for the angular scales retained in the analysis.

In practice, this approximation is implemented by applying the mask to the source estimates after the sparse-regularization step,
\begin{equation}
    \mathbf{S}^{M,(j)}_{n}(p)
    =
    M(p)\,
    \mathbf{S}^{(j)}_{{\rm thr},n}(p),
    \label{eq:jessnet_masked_sources}
\end{equation}
where $\mathbf{S}^{(j)}_{{\rm thr},n}$ is the regularized estimate of source component $n$ in angular window $j$. The same observed region is used when computing the MAD that sets the learnlet noise level, so that masked pixels do not affect the threshold calibration.

The convergence criterion is also evaluated only on the observed sky,
\begin{equation}
    \Delta_S
    =
    \frac{
    \left\|
    \left(\mathbf{S}^{M}\right)^{(t)}
    -
    \left(\mathbf{S}^{M}\right)^{(t-1)}
    \right\|_{\rm F}
    }{
    \left\|
    \left(\mathbf{S}^{M}\right)^{(t)}
    \right\|_{\rm F}
    },
    \label{eq:jessnet_convergence_mask}
\end{equation}
where $\|\cdot\|_{\rm F}$ denotes the Frobenius norm and the superscript $M$ denotes masked source estimates. This prevents excluded pixels from driving the stopping criterion.

Equation~\ref{eq:jessnet_approx_masked_model} should not be interpreted as a full correction for mask-induced mode coupling. Rather, it is a practical approximation that allows the reconstruction to be constrained to the observed sky while avoiding an explicit inversion of the mask-coupling matrix. The recovered H\,{\sc i} statistics are therefore evaluated on a conservative analysis region, away from the mask boundary where the approximation is least accurate. The validity of this treatment is assessed empirically in Sect.~\ref{sect:results} using the survey-footprint configuration.

\subsection{Algorithm}

The full JESSNet procedure combines multiscale angular-windowing, beam-aware source separation, spherical-wavelet learnlet regularization, and mask projection. The algorithm proceeds as follows:

\begin{enumerate}
    \item[\textbf{(i)}] \textit{Angular-window construction.}
    A set of angular windows $W_j(\ell)$ is defined over the retained multipole range. In this work, we use cosine-tapered windows normalized to satisfy Eq.~\ref{eq:jessnet_window_partition}.

    \item[\textbf{(ii)}] \textit{Channel selection.}
    For each angular window, the effective beam response $\overline{B}_{\nu}^{(j)}$ is computed using Eq.~\ref{eq:jessnet_weighted_beam_response}. The retained channel set $\mathcal{C}_j$ is then defined by Eq.~\ref{eq:jessnet_channel_selection}.

    \item[\textbf{(iii)}] \textit{Large-scale PCA cleaning.}
    The lowest angular window is cleaned using PCA. Before applying PCA, the channels are brought to a common angular response by smoothing them to the beam of the lowest-frequency channel. The reconstructed H\,{\sc i} residuals are retained for this window.

    \item[\textbf{(iv)}] \textit{JESSNet source separation.}
    For each remaining angular window, the windowed harmonic coefficients are passed to JESSNet together with the beam transfer functions of the retained channels. The mixing matrix is initialized from a PCA of the data reconvolved to a common effective resolution.

    \item[\textbf{(v)}] \textit{Source update.}
    For fixed $\mathbf{A}^{(j)}$, the source coefficients are estimated in harmonic space using the Tikhonov-regularized beam-aware update of Eq.~\ref{eq:jessnet_source_update}. We particularly follow the mixing-matrix-based strategy \cite[strategy \#3 from][]{carloni2021}.

    \item[\textbf{(vi)}] \textit{Spherical-wavelet learnlet regularization.}
    The source maps are transformed to map space, normalized on the observed sky region, decomposed into spherical-wavelet scales, and converted into aligned multiscale patch cubes. These cubes are passed through the trained learnlet operator $\mathcal{L}_{\theta}$, which returns denoised map patches. The patches are reassembled, unnormalized, and projected onto the observed sky.

    \item[\textbf{(vii)}] \textit{Mixing-matrix update.}
    For fixed regularized source estimates, $\mathbf{A}^{(j)}$ is updated by least squares using the detail scales of the data and sources. The columns of $\mathbf{A}^{(j)}$ are then normalized, and positivity can be imposed when required.

    \item[\textbf{(viii)}] \textit{Convergence and recombination.}
    The source and mixing-matrix updates are alternated until convergence, evaluated on the unmasked pixels. The cleaned H\,{\sc i} contributions from all angular windows are finally recombined according to Eq.~\ref{eq:jessnet_multiscale_recombination}.
\end{enumerate}

\subsection{Parameters used in the experiments}

The simulations in this work are analyzed at ${\rm N}_{\rm side}=256$, with spherical harmonic transforms computed up to $\ell_{\rm max}=3\,{\rm N}_{\rm side}$. The multiscale reconstruction is restricted to multipoles $\ell\leq\ell_{\rm up}$, with $\ell_{\rm up}={\rm N}_{\rm side}$. The angular windows are
constructed from the multipole boundaries $\ell=\{20,100\}$, and the cosine taper width is set to $\Delta\ell=10$. The channel-selection threshold is fixed to $\tau=0.1$, corresponding to a beam response equal to $10\%$ of its large-scale value.

The lowest angular window is cleaned with PCA using $N_s=4$. For the JESSNet windows, we use $N_s=5$ effective components. This choice is further explained in Appendix~\ref{app:ns_choice}. The source update uses the mixing-matrix-based Tikhonov regularization strategy of \citet{carloni2021}, corresponding to their Strategy~\#3. The regularization parameter is set to
$c_{\rm wu}=10^{-2}$. The minimum number of warm-up iterations is set to $N_{\rm wu}=50$.

Convergence is monitored through the relative change of the source estimates on the observed sky region. The warm-up stage is stopped when $\Delta_S \leq 10^{-2}$ after the minimum number of warm-up iterations, or after the maximum allowed number of warm-up iterations is reached. The refinement stage is stopped when $\Delta_S \leq 10^{-4}$, with a minimum of 25 refinement iterations. A final source-only refinement is performed with the mixing matrix fixed, using the same convergence tolerance $10^{-4}$.

The spherical-wavelet learnlet uses $L=5$ spherical scales, corresponding to one coarse scale and four detail scales. The learnlet architecture uses 64 filters, a kernel size of 5, and hard thresholding. For the maps analyzed at ${\rm N}_{\rm side}=256$, each spherical-wavelet coefficient map is partitioned into 192 local HEALPix patches of size $64\times64$ pixels before being passed to the learnlet. The learnlet weights are trained on simulated foreground maps with varying spectral behavior. These parameter choices are kept fixed for the full-sky and survey-footprint analyses presented in Sect.~\ref{sect:results}.

For reference, on a single 64-core node, the fiducial multiscale JESSNet reconstruction with $N_s=5$ requires approximately $20$ min for the full-sky analysis and $29$ min for the survey-footprint analysis, for $N_{\rm side}=256$ and 400 frequency channels. The footprint considered here covers approximately $17\%$ of the sky. Despite its smaller sky coverage, the footprint analysis is somewhat more computationally expensive, reflecting the additional iterations needed to converge to a solution near the mask boundary.

\section{Results}
\label{sect:results}

We assess the quality of the reconstructed H\,{\sc i} signal through its angular and frequency power spectra. We present results for the multiscale foreground-cleaning strategy described in Sect.~\ref{sect:jessnet}, which combines PCA on the largest angular scales with JESSNet in the higher angular windows. In the remainder of this section, we refer to this multiscale configuration simply as JESSNet. For comparison, we also show the reconstruction obtained by applying the same JESSNet source-separation procedure over a single angular range, without the multiscale angular-window treatment. We refer to this baseline as monoscale-JESSNet. We first evaluate both approaches on full-sky simulations and then on a synthetic analysis footprint designed to test incomplete sky coverage.

\subsection{Full sky}

\begin{figure*}[t!]
    \centering
    \begin{minipage}{0.49\textwidth}
        \centering
        \includegraphics[width=\textwidth]{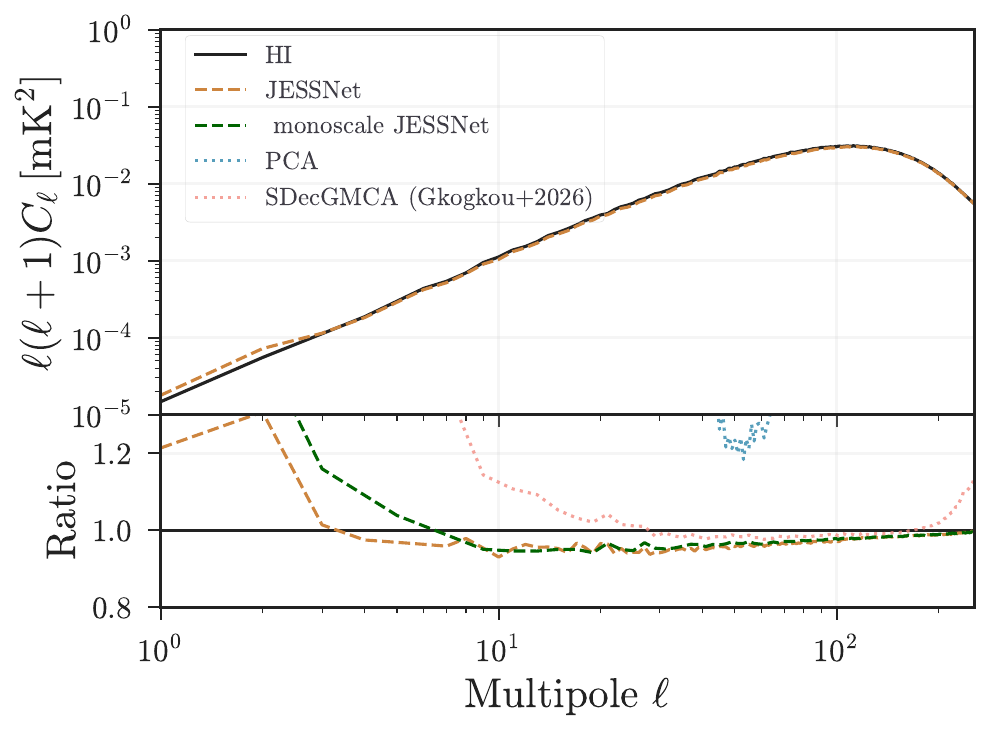}
    \end{minipage}
    \hfill
    \begin{minipage}{0.49\textwidth}
        \centering
        \includegraphics[width=\textwidth]{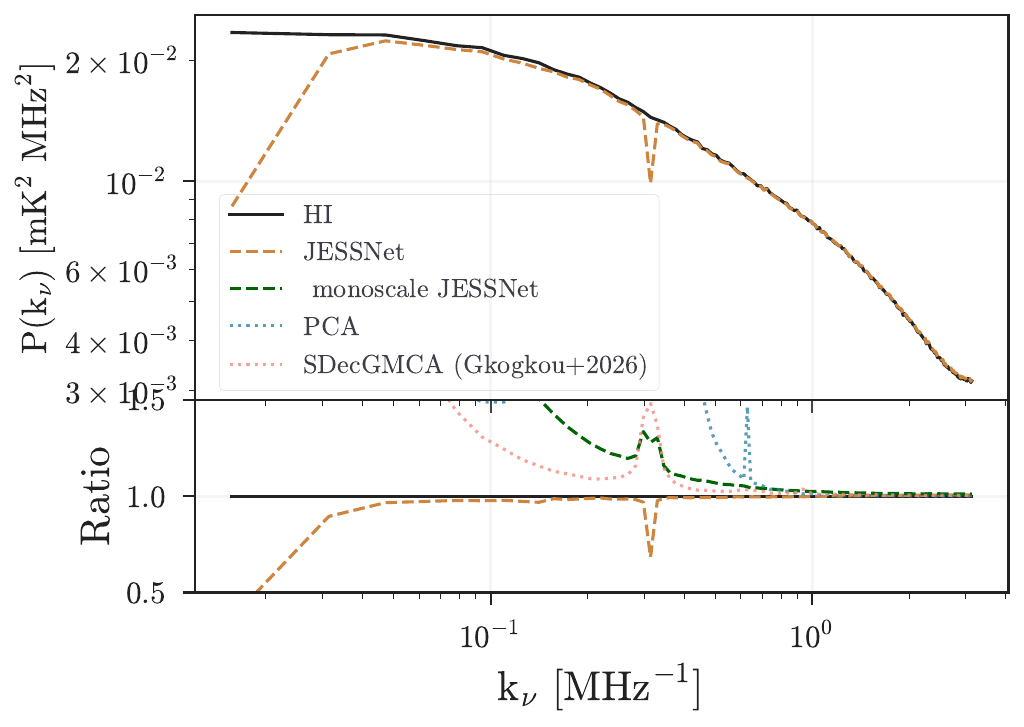}
    \end{minipage}
    \caption{Recovery of the H\,{\sc i} power spectra in the full-sky analysis. Left: angular power spectrum. Right: frequency power spectrum. In both panels, the black solid curve shows the input H\,{\sc i} signal, the orange dashed curve corresponds to the JESSNet reconstruction, and the green dashed curve shows the monoscale-JESSNet reconstruction. The lower subpanels display the ratio of each reconstructed spectrum to the input H\,{\sc i} spectrum; for reference, the same subpanels also show the ratios obtained with PCA and with SDecGMCA \citep{gkogkou2026}. For the standalone PCA comparison, five principal components are removed, matching the foreground-subspace dimensionality $N_s=5$ adopted for JESSNet.}
    \label{fig:results_fullsky}
\end{figure*}

This configuration isolates the effect of the multiscale cleaning strategy and the spherical-wavelet learnlet regularization from complications associated with incomplete sky coverage. Fig.~\ref{fig:results_fullsky} shows the recovered angular and frequency power spectra.

The JESSNet reconstruction closely follows the input H\,{\sc i} angular power spectrum over the multipole range retained in the analysis. The largest deviations occur at the lowest multipoles, where foreground cleaning is intrinsically more difficult because both the foregrounds and the cosmological signal are dominated by large-scale modes. At intermediate and small angular scales, the reconstructed spectrum remains close to the input H\,{\sc i} spectrum, with the ratio approaching unity over most of the range. By contrast, the monoscale-JESSNet reconstruction shows a stronger excess of power at low and intermediate multipoles, indicating that a single effective mixing model is less able to describe the foreground contamination across the full angular range.

A similar trend is observed in the frequency power spectrum. JESSNet reproduces the overall shape and amplitude of the input H\,{\sc i} spectrum over most of the probed line-of-sight modes. The main discrepancy appears at the lowest $k_\nu$ values, where the separation is limited by the degeneracy between spectrally smooth foregrounds and the largest radial H\,{\sc i} modes. A localized deviation is also visible at intermediate $k_\nu$, associated with the oscillatory chromatic structure of the instrumental beam. The main feature occurs at $k_\nu=2\pi/T\simeq0.314\,{\rm MHz}^{-1}$ for the adopted $T=20\,{\rm MHz}$ beam oscillation. A weaker secondary feature near $k_\nu\simeq0.63\,{\rm MHz}^{-1}$ is consistent with the second harmonic generated by the quadratic dependence of the beam response on its width. This feature remains confined to a narrow range of modes and is substantially reduced relative to the monoscale-JESSNet reconstruction.

This behavior is consistent with the systematic effects identified in previous foreground-cleaning studies. In \citet{gkogkou2026}, preprocessing of bright spectral peaks was shown to mitigate the broadband bias in the reconstructed frequency power spectrum, while SDecGMCA substantially suppressed the localized spurious feature around $k_\nu \sim 0.3$ associated with beam--foreground coupling and previously reported by \citet{spinelli2021}. The wavelet-based preprocessing adopted here similarly avoids the broadband bias, while JESSNet further suppresses the residual beam-induced feature compared with SDecGMCA, with the reconstructed-to-input ratio remaining closer to unity around the affected modes.

Fig.~\ref{fig:results_fullsky} additionally shows, for reference, the ratios obtained with PCA in the same full-sky configuration. For the five-mode configuration shown here, PCA exhibits a larger departure from unity than JESSNet, most notably around the modes affected by the oscillatory chromatic structure of the beam. This is consistent with the fact that PCA does not explicitly model the frequency dependence of the beam response, as already identified in \citet{gkogkou2026}. The comparison with SDecGMCA is more nuanced. Over part of the multipole and $k_\nu$ range, the monoscale-JESSNet ratio is slightly farther from unity than the SDecGMCA ratio. This difference should not be over-interpreted. The SDecGMCA results of \citet{gkogkou2026} benefited from a case-specific optimization of the regularization hyperparameters, informed by knowledge of the true H\,{\sc i} signal, an option unavailable for real data and not used in JESSNet either. JESSNet's hyperparameters, in both its monoscale and multiscale configurations, are instead fixed a priori and require no advance knowledge of the ground truth. Despite this more conservative setting, the multiscale JESSNet reconstruction remains closer to unity over most of the probed angular and radial scales.

Overall, the full-sky analysis demonstrates the benefit of the multiscale formulation. By allowing both the effective mixing matrix and the retained channel set to vary with angular scale, JESSNet avoids forcing poorly resolved channels to contribute to small-scale modes while still exploiting all available frequency information on large angular scales. This leads to a more stable foreground subtraction and a more accurate recovery of the H\,{\sc i} power spectra than the monoscale-JESSNet configuration.

\subsection{Synthetic analysis footprint}

Real observations do not generally provide full-sky maps, but instead cover only a fraction of the sky. This motivates testing JESSNet under incomplete sky coverage, where survey boundaries or data-quality masks restrict the region available for component separation.

\begin{figure}[t!]
    \centering
    \includegraphics[width=0.7\linewidth]{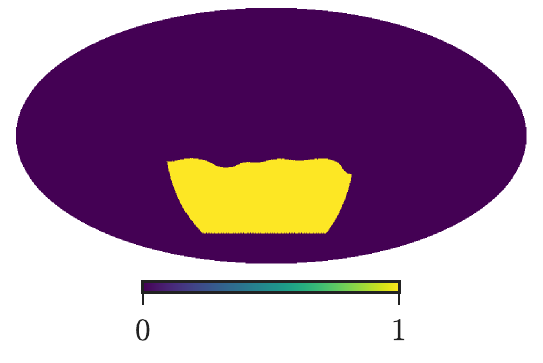}
    \caption{Synthetic analysis footprint used to test the JESSNet reconstruction under incomplete sky coverage. Pixels with value 1 are retained in the analysis, while pixels with value 0 are excluded.}
    \label{fig:footprint_mask}
\end{figure}

To assess the behavior of the method in this setting, we adopt a synthetic analysis footprint designed solely as a controlled test of the mask-constrained reconstruction. The footprint is not intended to reproduce the geometry of any specific survey, but simply defines a localized observed region on the sky. The resulting binary mask is shown in Fig.~\ref{fig:footprint_mask}, with $M(p)=1$ for pixels retained in the analysis and $M(p)=0$ elsewhere.

For the masked-sky analysis, we restrict the comparison of angular power spectra to the multipole range $7 < \ell < 250$. The lowest multipoles are excluded because the pseudo-$C_\ell$ estimates are strongly affected by mode coupling induced by the finite footprint, while the upper limit corresponds to the maximum multipole retained in the multiscale reconstruction. We therefore quote results only in the range where the reconstruction is not dominated by footprint-induced leakage or by the explicit harmonic truncation.

\begin{figure*}[h!]
    \centering
    \begin{minipage}{0.49\textwidth}
        \centering
        \includegraphics[width=\textwidth]{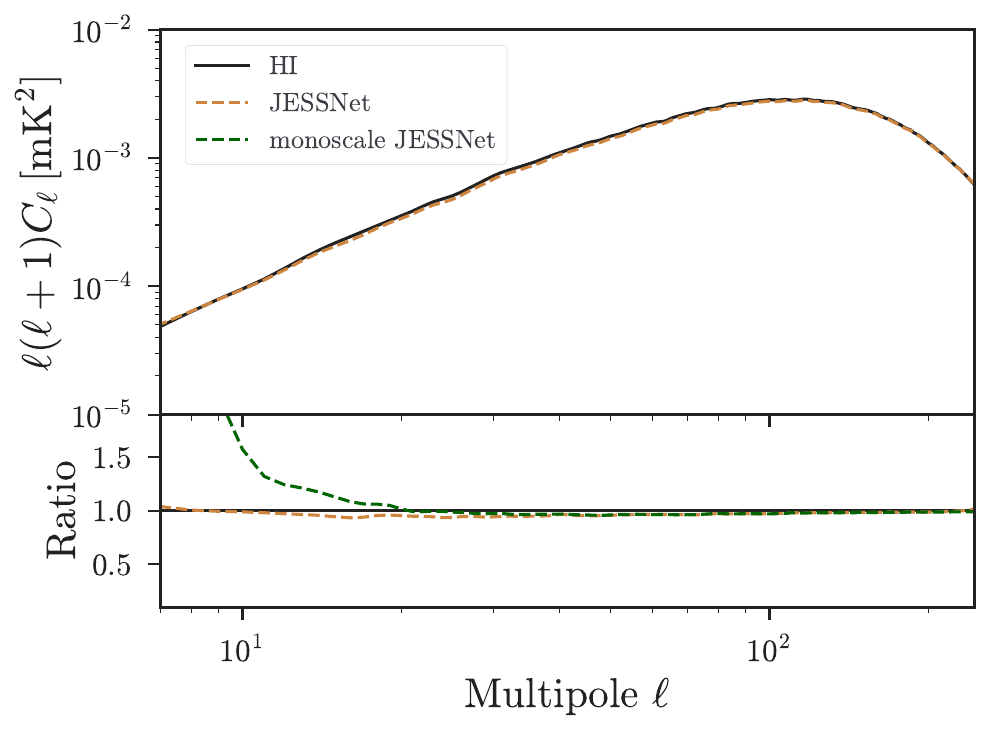}
    \end{minipage}
    \hfill
    \begin{minipage}{0.49\textwidth}
        \centering
        \includegraphics[width=\textwidth]{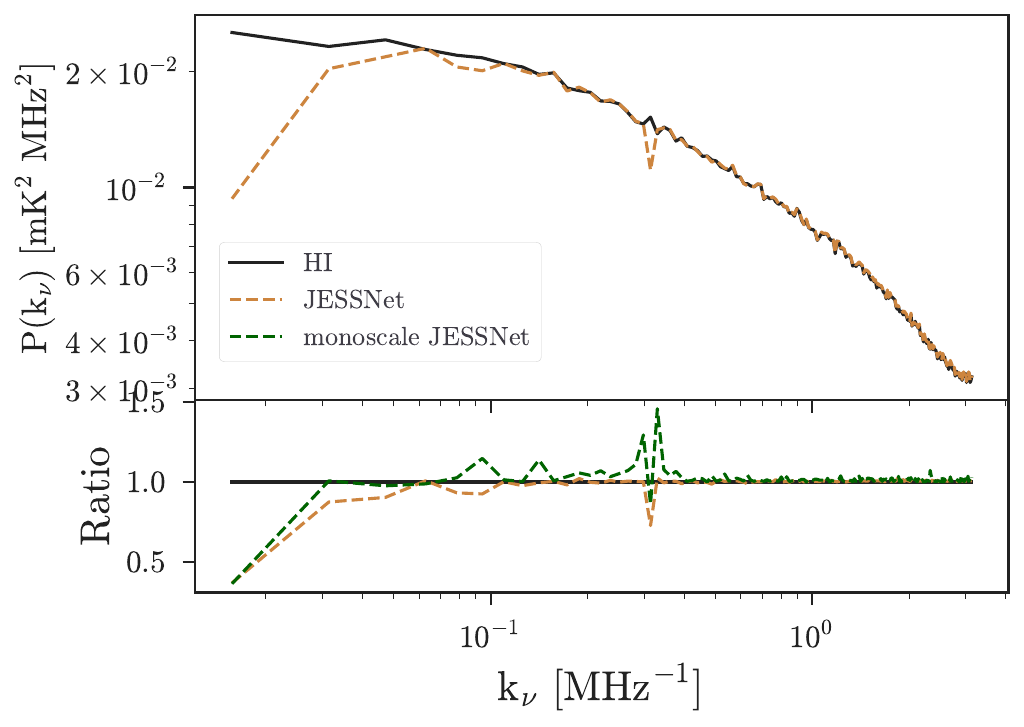}
    \end{minipage}
    \caption{Recovery of the H\,{\sc i} power spectra for the synthetic-footprint analysis. Left: angular power spectrum. Right: frequency power spectrum. In both panels, the black solid curve shows the input H\,{\sc i} signal, the orange dashed curve corresponds to the JESSNet reconstruction, and the green dotted curve shows the monoscale-JESSNet reconstruction. The lower subpanels display the ratio of each reconstructed spectrum to the input H\,{\sc i} spectrum.}
    \label{fig:results_footprint}
\end{figure*}

The recovered spectra for the synthetic-footprint configuration are shown in Fig.~\ref{fig:results_footprint}. Despite the substantially reduced sky coverage, the JESSNet reconstruction remains in close agreement with the input H\,{\sc i} signal. In the angular power spectrum, the reconstructed curve follows the true spectrum throughout the multipole range considered, with the ratio remaining close to unity over most of the range. The monoscale-JESSNet reconstruction shows a larger excess at the lowest multipoles of the retained interval, demonstrating that the multiscale treatment remains beneficial when the sky coverage is incomplete.

The frequency power spectrum also remains well recovered over most of the probed $k_\nu$ range. As in the full-sky case, the largest departure occurs at the lowest line-of-sight modes, where foreground cleaning is limited by the intrinsic spectral degeneracy between the foregrounds and the cosmological signal. A localized feature associated with the oscillatory beam response is again visible at intermediate $k_\nu$, but the JESSNet reconstruction remains close to the input spectrum over the rest of the range. The monoscale-JESSNet reconstruction exhibits stronger localized deviations, confirming that the angular-window decomposition improves the stability of the separation.

These results demonstrate that the mask-constrained implementation of JESSNet remains effective under restricted sky coverage. Although the footprint introduces mode coupling and reduces the amount of angular information available, the reconstructed H\,{\sc i} statistics remain well preserved over the multipole and frequency ranges used in the analysis. This supports the approximation introduced in Sect.~\ref{sect:jessnet}, in which the reconstruction is constrained to the observed sky and the recovered statistics are evaluated away from the mask boundaries.

\subsection{Discussion}
\label{subsect:results_discussion}

The full-sky and survey-footprint analyses show that JESSNet can recover the main angular and radial H\,{\sc i} statistics in the presence of a chromatic beam and incomplete sky coverage. Beyond these controlled tests, however, several practical aspects are relevant when considering the application of the method to observational data.

A first point concerns the comparison with PCA. We verified that increasing the PCA rank can improve the reconstruction in one diagnostic without yielding a comparable improvement in the other. In particular, a rank that performs well for $C_\ell$ does not necessarily provide the best recovery of $P(k_\nu)$, and vice versa. This reflects the usual compromise between residual foreground contamination and H\,{\sc i} signal loss. PCA therefore remains an effective and computationally inexpensive baseline and, with an appropriate choice of rank, can provide a reasonably accurate reconstruction in the controlled simulations considered here. JESSNet is substantially more computationally demanding, but its additional complexity provides a framework in which the instrumental beam is modeled explicitly rather than absorbed implicitly into the foreground subspace. This becomes particularly relevant when chromatic beam effects limit the separation and also provides a natural starting point for extending the method to more complex instrumental responses.

The usefulness of this beam-aware formulation naturally depends on the accuracy of the assumed instrumental response. The present implementation adopts an axisymmetric, frequency-dependent beam model. The sensitivity tests presented in Appendix~\ref{ap:beam_uncertainty} show that moderate uncertainties in the assumed beam width have little impact on the recovered power spectra, whereas larger mismatches progressively degrade the reconstruction. A significantly asymmetric beam would require extending the JESSNet forward model beyond the scalar harmonic transfer function adopted here in order to account for the associated mode coupling.

Finally, the present analysis does not explicitly include residual radio-frequency interference (RFI) or other non-astrophysical spectral systematics. Such contamination may not follow the beam-convolved mixture model assumed by JESSNet and could therefore affect the PCA initialization, the scale-dependent mixing matrices, or the learned sparse regularization. In applications to real observations, RFI flagging or downweighting should therefore be applied before foreground cleaning, with data-quality information incorporated into the retained channel sets. We nevertheless tested the effect of RFI-like channel flagging by removing combinations of contiguous and isolated frequency channels. The angular reconstruction remains comparatively robust even when up to $40\%$ of the channels are unavailable: the RMS deviation of $C_\ell^{\rm rec}/C_\ell^{\rm HI}$ remains at the few-percent level, reaching approximately $5\%$ in both the full-sky and footprint cases. The radial power spectrum is substantially more sensitive, with the RMS deviation rising to approximately $50\%$ for the same missing-channel fraction. This behavior reflects the need to reconstruct a continuous frequency axis across increasingly large spectral gaps and indicates that improved gap-filling strategies or frequency-domain estimators will be particularly important for applications to heavily flagged  observational data.

\section{Conclusions}
\label{sect:conclusions}

In this work, we introduced JESSNet, a multiscale foreground-cleaning framework for H\,{\sc i} 21 cm intensity mapping over the redshift range
$z\simeq0.09\text{--}0.58$. JESSNet preserves the harmonic-space treatment of the frequency-dependent beam, but extends SDecGMCA in three main directions. First, the reconstruction is performed in angular windows, with an effective mixing matrix and a retained channel set adapted to each angular scale. Second, the fixed starlet thresholding step is replaced by a spherical-wavelet learnlet sparse regularization operator. Third, the reconstruction is constrained to the observed sky region, enabling applications to masks and survey footprints. Throughout, the reconstructed components should be understood as effective latent quantities that span the foreground subspace, rather than as physical estimates of individual emission mechanisms.

The multiscale formulation is motivated by the fact that different frequency channels carry reliable angular information only over a limited range of scales when the beam is chromatic. JESSNet exploits this by allowing the effective mixing matrix and the retained channel set to adapt to each angular scale, avoiding the noise amplification that a single, scale-independent model would introduce.

We tested the method on simulated single-dish H\,{\sc i} intensity mapping observations covering $900\text{--}1300\,{\rm MHz}$. The simulations include a cosmological H\,{\sc i} signal generated with \texttt{CoLoRe}, Galactic and extragalactic foregrounds, thermal noise, and an oscillating in frequency MeerKAT-inspired beam. The largest angular scales are cleaned with PCA after beam homogenization, while JESSNet is applied at smaller angular scales using cosine-tapered multipole bands and a scale-dependent channel selection based on the beam response.

On full-sky simulations, the JESSNet reconstruction recovers both the angular and frequency power spectra of the input H\,{\sc i} signal with good accuracy over the scales retained in the analysis. The main residual deviations occur at the largest angular scales, the lowest line-of-sight modes, and in a localized range associated with the oscillatory beam response. Compared to applying monoscale-JESSNet or SDecGMCA, the multiscale strategy reduces the excess power at low and intermediate multipoles and improves the stability of the reconstruction around beam-induced spectral features. Notably, this improvement over SDecGMCA is obtained without any case-specific hyperparameter optimization informed by the ground truth, unlike the SDecGMCA results of \citet{gkogkou2026}, underlining that the gain follows from the multiscale formulation itself.

We also applied the method to a synthetic analysis footprint designed to test the reconstruction under incomplete sky coverage. This test demonstrates that the mask-constrained implementation can operate under restricted sky coverage while preserving the main H\,{\sc i} statistics. Although the method does not explicitly invert the full mask-induced mode-coupling matrix, evaluating the reconstructed signal on a conservative analysis region away from the mask boundaries yields angular and frequency power spectra that remain close to the input H\,{\sc i} spectra over the ranges considered.

The realistic oscillating beam model and mask-constrained formulation considered here address two important complications expected in ongoing and upcoming single-dish H\,{\sc i} intensity-mapping surveys, such as MeerKLASS and SKA-MID. Extending the validation to additional real-data systematics, including RFI and imperfect instrumental characterization, will be an important step toward application to observational data. The code and simulated data used in this work are publicly released, allowing the community to reproduce the results presented here and to adapt the method to other instruments, beam models, and survey footprints.

\begin{acknowledgements}
      This work was funded by the TITAN ERA Chair project (contract no. 101086741) within the Horizon Europe Framework Program of the European Commission, the French Agence Nationale de la Recherche (ANR-22-CE31-0014-01 TOSCA), and the French Programme National de Cosmologie et Galaxies (PNCG project CIMES).
\end{acknowledgements}

\bibliographystyle{aa}
\bibliography{aa}

\appendix
\raggedbottom

\section{Sensitivity to beam-model uncertainties}
\label{ap:beam_uncertainty}

The JESSNet reconstruction requires an estimate of the frequency-dependent beam response for both the beam-aware source update and the scale-dependent channel selection. To assess the impact of imperfect beam knowledge, we keep the simulated observations fixed, convolved with the fiducial beam described in Sect.~\ref{subsect:telescope_beams}, and perturb only the beam model supplied to JESSNet. Specifically, the assumed FWHM is rescaled by $\pm5\%$, $\pm10\%$, and $\pm20\%$, in addition to the fiducial case.

\begin{figure*}[t!]
    \centering

    \begin{minipage}{0.49\textwidth}
        \centering
        \includegraphics[width=\textwidth]{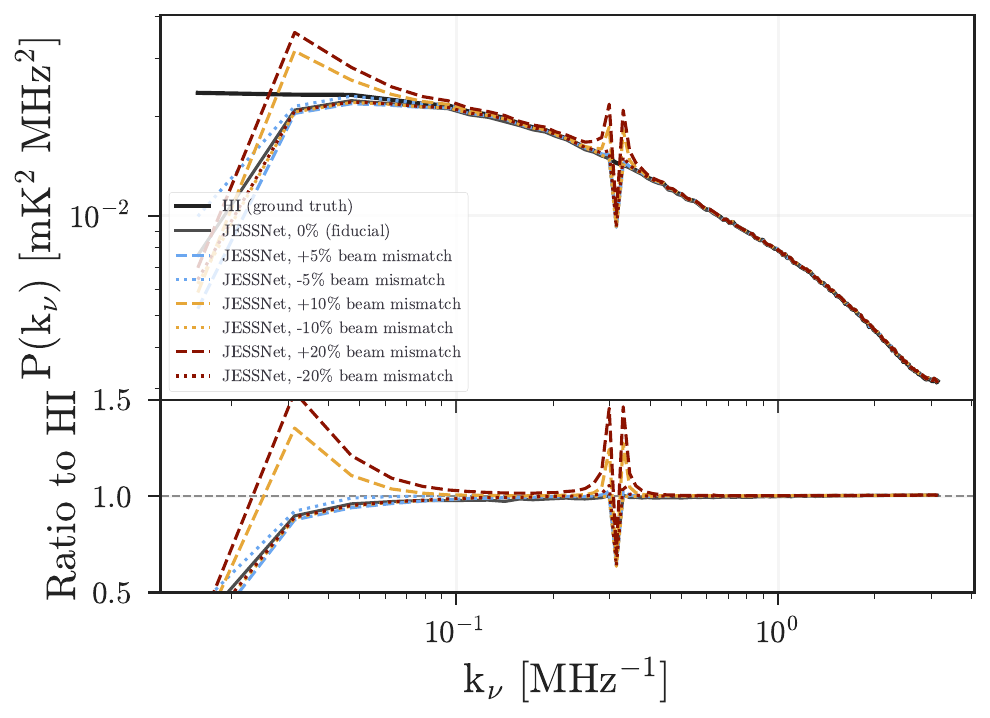}
    \end{minipage}
    \hfill
    \begin{minipage}{0.49\textwidth}
        \centering
        \includegraphics[width=\textwidth]{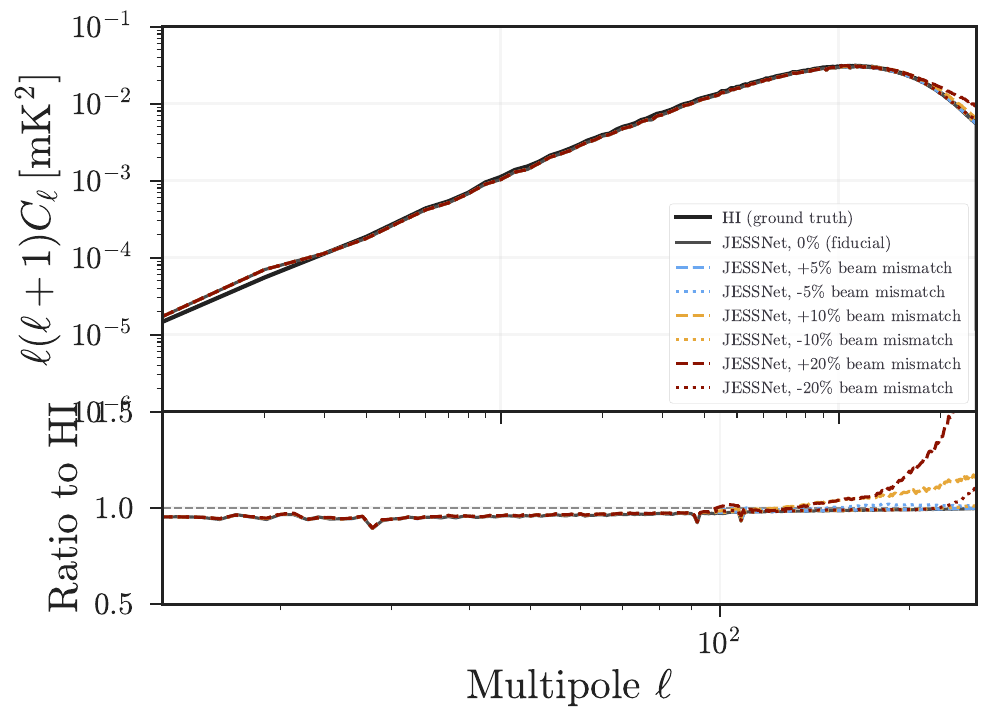}
    \end{minipage}

    \vspace{0.4cm}

    \begin{minipage}{0.49\textwidth}
        \centering
        \includegraphics[width=\textwidth]{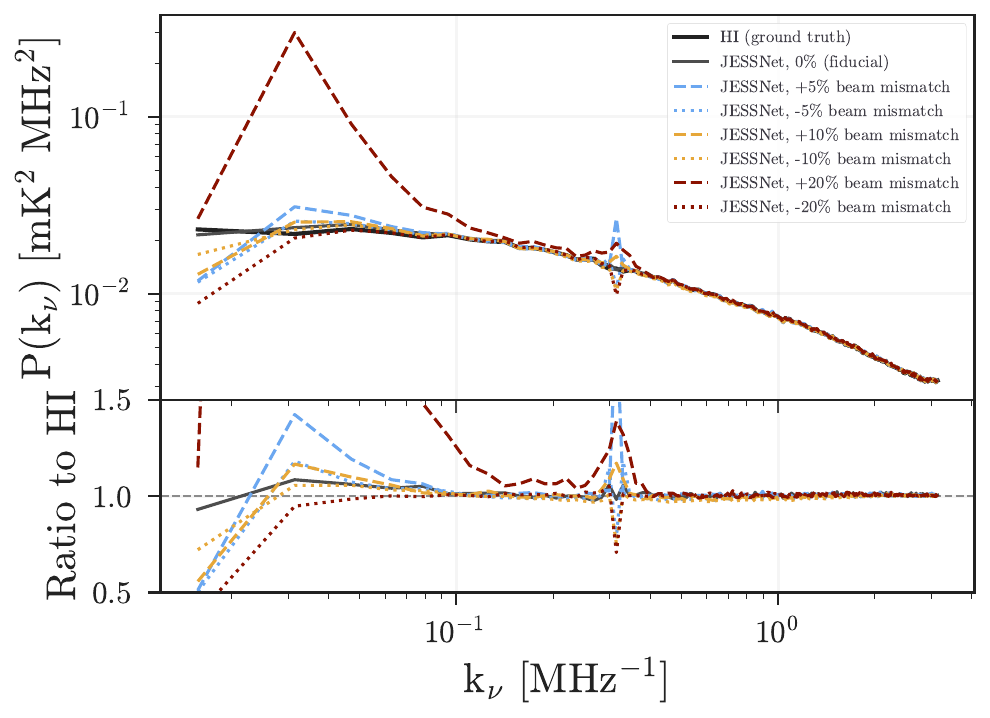}
    \end{minipage}
    \hfill
    \begin{minipage}{0.49\textwidth}
        \centering
        \includegraphics[width=\textwidth]{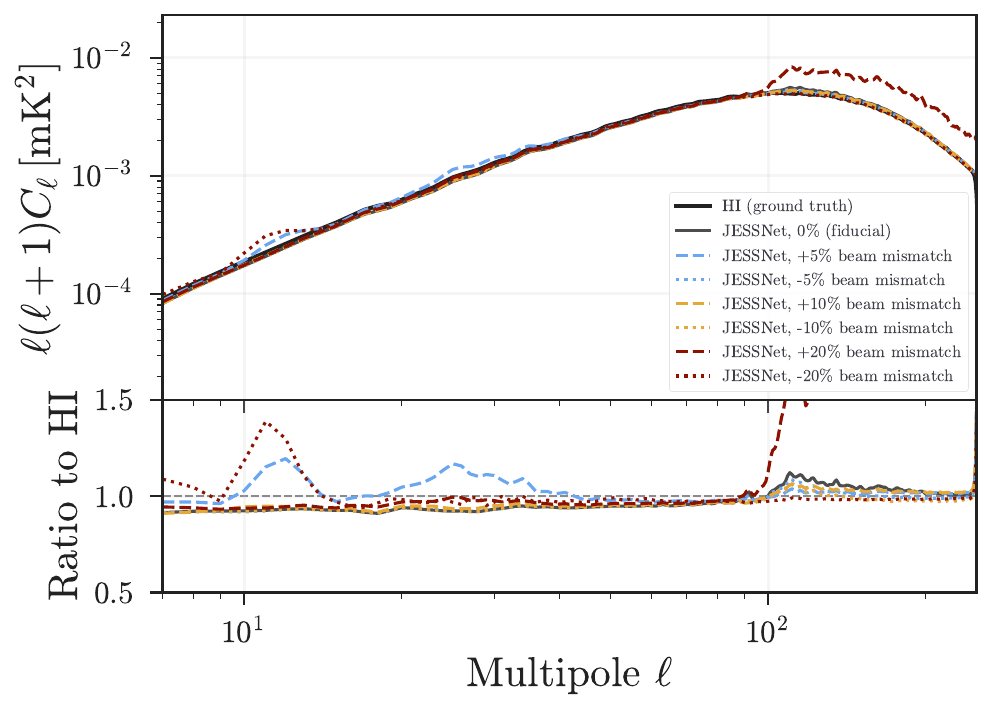}
    \end{minipage}

    \caption{
    Sensitivity of the JESSNet reconstruction to uncertainties in the assumed beam model. Top row: full-sky configuration. Bottom row: survey-footprint configuration. The left panels show the recovered frequency power spectrum $P(k_\nu)$ and the right panels show the angular power spectrum $C_\ell$, for different fractional mismatches between the true beam FWHM used to generate the observations and the beam model supplied to JESSNet.}
    \label{fig:beam_uncertainty}
\end{figure*}

Figure~\ref{fig:beam_uncertainty} shows the resulting angular and frequency power spectra for the full-sky and survey-footprint configurations. For each case, we quantify the reconstruction error through the RMS deviation of $C_\ell^{\rm rec}/C_\ell^{\rm HI}$ and $P(k_\nu)^{\rm rec}/P(k_\nu)^{\rm HI}$ from unity, evaluated over $13\leq\ell\leq250$ and $k_\nu>0.05\,{\rm MHz}^{-1}$, respectively. These ranges exclude the large-scale regime in which foreground--H\,{\sc i} degeneracy already limits the reconstruction independently of the beam model.

The results indicate that the reconstruction is stable against beam-FWHM errors at the few-percent level. In the full-sky case, for example, the RMS angular-spectrum error remains approximately $2.6$--$2.9\%$ for mismatches up to $\pm5\%$, compared with $2.9\%$ for the fiducial beam. Larger beam errors produce progressively stronger deviations, particularly at high multipoles and around the spectral feature associated with the oscillatory beam response. A similar qualitative behavior is found for the survey-footprint analysis. These results indicate that modest beam-model uncertainties do not strongly affect the reconstruction, while sufficiently large errors in the assumed beam response propagate into the recovered H\,{\sc i} statistics.

The present implementation assumes an axisymmetric beam. A significantly asymmetric response would require extending the forward model to a more general beam operator that accounts for the associated harmonic-mode coupling.

\section{Choice of the JESSNet foreground-subspace dimension}
\label{app:ns_choice}

The JESSNet reconstruction requires specifying the number of effective foreground components, $N_s$. To assess the sensitivity of the reconstruction to this choice, we repeat the complete analysis for $N_s$ in the range $[3,9]$ in both the full-sky and survey-footprint configurations, while keeping all other parameters fixed.

\begin{figure*}[t!]
    \centering

    \begin{minipage}{0.49\textwidth}
        \centering
        \includegraphics[width=\textwidth]{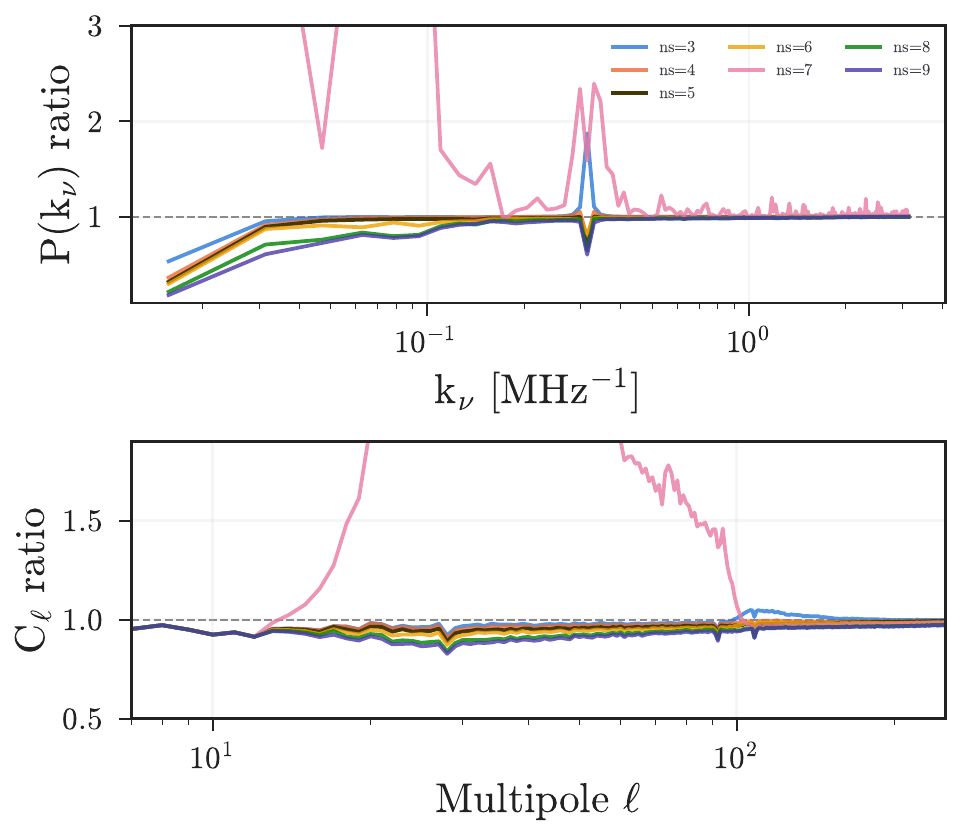}
    \end{minipage}
    \hfill
    \begin{minipage}{0.49\textwidth}
        \centering
        \includegraphics[width=\textwidth]{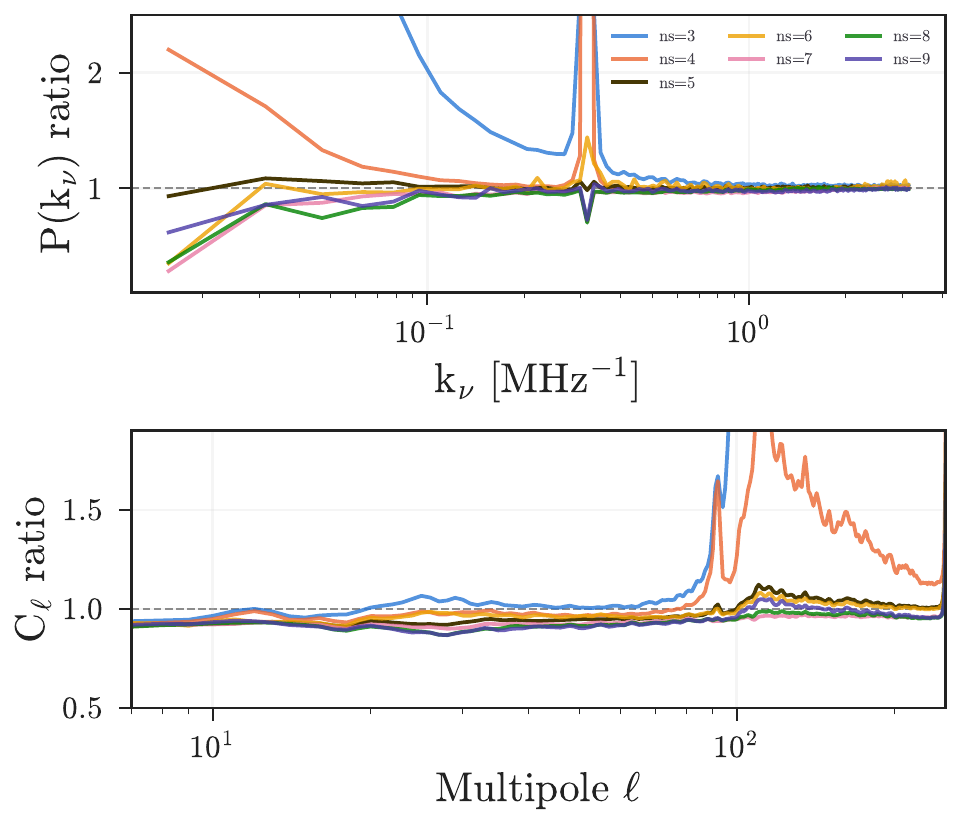}
    \end{minipage}

    \caption{
    Sensitivity of the JESSNet reconstruction to the number of effective
    foreground components $N_s$. Left: full-sky configuration. Right:
    survey-footprint configuration. The upper panels show the ratio of the
    reconstructed frequency power spectrum to the input H\,{\sc i} spectrum,
    while the lower panels show the corresponding angular power-spectrum
    ratio. Results are shown for $N_s=3,\ldots,9$, with all other JESSNet
    parameters kept fixed.
    }
    \label{fig:ns_sweep}
\end{figure*}

Figure~\ref{fig:ns_sweep} shows the ratios of the reconstructed frequency and angular power spectra to the input H\,{\sc i} spectra. The dependence on $N_s$ is most pronounced in the survey-footprint case, where $N_s=3$ and $4$ leave substantial residual contamination, while $N_s=5$ is the smallest tested value for which both diagnostics become stable. Increasing $N_s$ further does not lead to a systematic improvement in the reconstruction, and some larger values can introduce numerical instabilities, as seen for $N_s=7$ in the full-sky case.

Based on this test, we adopt $N_s=5$ throughout the analysis. This is the smallest tested value that avoids the clear underfitting regime in both the full-sky and survey-footprint configurations. Larger values can also provide accurate reconstructions, but do not yield a systematic improvement. We therefore retain a single common value rather than optimizing $N_s$ separately for each survey geometry.

\section{Public code}
\label{ap:public_code}

The code used to run the JESSNet foreground-cleaning pipeline is publicly available \footnote{\url{https://github.com/Agkogkou/JESSNet}}. The repository implements the JESSNet solver described in Sect.~\ref{sect:jessnet}, the windowed cleaning pipeline, and the spherical-wavelet learnlet architecture, together with the script used to train the learnlet and the script used to reproduce the full-sky and synthetic-footprint results of Sect.~\ref{sect:results}. The code requires Python~3.9 or later and standard scientific packages ({\tt numpy}, {\tt scipy}, {\tt healpy}, {\tt PyTorch}).

The large binary files required to reproduce the numerical results are publicly available on Zenodo \footnote{\url{https://doi.org/10.5281/zenodo.20341922}}. These include the simulated H{\sc i} intensity-mapping data cube, the masks used for the survey-footprint analysis, and the trained learnlet weights.

\end{document}